\documentclass[10pt
]{article}
\usepackage{geometry}
\usepackage{tabularx}
\usepackage{xcolor}
\usepackage{graphicx}
\usepackage{dcolumn}
\usepackage{bm}
\usepackage{enumitem}
\usepackage{authblk}
\usepackage{amsthm}
\usepackage{amsmath}
\usepackage{amsfonts}
\usepackage{nameref}
\usepackage{mathtools}

\newtheoremstyle{break}
  {\topsep}{\topsep}%
  {\itshape}{}%
  {\bfseries}{}%
  {\newline}{}%
\theoremstyle{break}
\newtheorem{theorem}{Theorem}
\newtheorem{lemma}{Lemma}

\theoremstyle{remark}

\usepackage{chngcntr}

\usepackage{tikz,xcolor}
\usepackage[hypertexnames=false]{hyperref}
\definecolor{lime}{HTML}{A6CE39}
\DeclareRobustCommand{\orcidicon}{
	\begin{tikzpicture}
	\draw[lime, fill=lime] (0,0) 
	circle [radius=0.16] 
	node[white] {{\fontfamily{qag}\selectfont \tiny ID}};
	\draw[white, fill=white] (-0.0625,0.095) 
	circle [radius=0.007];
	\end{tikzpicture}
	\hspace{-2mm}
}

\foreach \x in {A, ..., Z}{\expandafter\xdef\csname orcid\x\endcsname{\noexpand\href{https://orcid.org/\csname orcidauthor\x\endcsname}
			{\noexpand\orcidicon}}
}

\providecommand{\keywords}[1]{\textit{Keywords:} #1}

\begin{document}

\title{Nonparametric estimation of the preferential attachment function from one network snapshot}

\author[1]{Thong Pham\orcidA{}\footnote{Corresponding author. Email: thong.pham@riken.jp}}
\author[2]{Paul Sheridan\orcidB{}}
\author[1,3]{Hidetoshi Shimodaira\orcidC{}}

\affil[1]{RIKEN Center for AIP}
\affil[2]{Tupac Bio, Inc.}
\affil[3]{Kyoto University}
\maketitle

\begin{abstract}
{Preferential attachment is commonly invoked to explain the emergence of those heavy-tailed degree distributions characteristic of growing network representations of divers{e} real-world phenomena. Experimentally confirming this hypothesis in real-world growing networks is an important frontier in network science research. Conventional preferential attachment estimation methods require that a growing network be observed across at least two snapshots in time. Numerous publicly available growing network datasets are, however, only available as single snapshots, leaving the applied network scientist with no means of measuring preferential attachment in these cases. We propose a nonparametric method, called PAFit-oneshot, for estimating preferential attachment in a growing network from one snapshot. PAFit-oneshot corrects for a previously unnoticed bias that arises when estimating preferential attachment values only for degrees observed in the single snapshot. Our work provides a means of measuring preferential attachment in a large number of publicly available one-snapshot networks. As a demonstration, we estimated preferential attachment in three such networks, and found sublinear preferential attachment in all cases. PAFit-oneshot is implemented in the \textsf{R} package \texttt{PAFit}.} 
\end{abstract}
\maketitle

\section{Introduction}
\label{sec: intro}
{Preferential attachment (PA) has been put forward as an intuitive explanation for the }surprising universality of heavy-tail degree distributions in both human-made and natural networks in {diverse} domains, {including} technology, biology, society, and linguistics~\cite{clauset,perc,power_law_finite_size_effect}. Historically, the discussion about heavy-tail degree distributions in the literature had centered around power-laws~\cite{barabasi-albert}, until it was pointed out in more recent times that heavy-tails can take on a rich variety of forms {(e.g., stretched-exponential, log-normal, power-law with a cut-off, etc.)}~\cite{clauset,power_law_myth, clauset_powerlaw_2,power_law_finite_size_effect}. {According to the PA mechanism, the probability a node with degree $k$ acquires a new edge is proportional to $A_k \ge 0$, the ``attachment'' value of degree $k$ nodes. The function $A_k$ is often called \emph{attachment kernel} or \emph{attachment function}. When $A_k$ is equal to $k + c$ for some constant $c \ge 0$, we have the celebrated linear preferential attachment function. It is the first PA function investigated by the complex network community and it is this form that leads to a power-law degree distribution~\cite{barabasi-albert}. Different functional forms of $A_k$ do, however, generate different types of heavy-tail degree-distributions~\cite{krapi2}.}

{It turns out that various well-known self-reinforcing mechanisms in a wide range of fields can be interpreted as some form of PA.} The first instance of such mechanisms arose in biology when the Yule process was proposed to describe the evolution of biological species~\cite{yule}. Simon {utilized} the same concept to explain various phenomena in biology, sociology, and economics~\cite{simon}. Merton coined the now-famous term ``Matthew effect'' when he used PA to explain the discrepancies in recognition between famous scientists and {lesser-known} ones~\cite{Merton56}. Price~\cite{price2,price1} used the phrase ``cumulative advantage'' when {he employed PA to elucidate various phenomena in scientometrics such as} Lotka's law in scientific productivity~\cite{lotka_law} and Bradford's law in journal use~\cite{bradford_law}. PA was also recently used in describing the herding effect in financial markets~\cite{herding_1}. Estimating PA from {the empirical growing network data is, therefore, }important not only for verifying various assumptions about how heavy-tail degree distributions arise, but also for providing new insights on the aforementioned related phenomena in various fields.

{We consider t}he problem of estimating $A_k$ {from empirical data that does not contain any information about the growth process of the network.} Consider a {growing} network that grows from time{-step} $t =1$ to $T$ {and d}enote its snapshot at time{-step} $t$ by $G_t$. {Traditionally, the estimation of $A_k$ is often considered when the growth process of the network can be observed at at least two time-steps~\cite{newman2001clustering,jeong,pham2}}. However, what if we cannot observe anything about the growth process and have to {content} ourselves with only {the one} snapshot $G_T$? Is it possible to recover $A_k$, a function that governs the growth process, without observing the growth process itself? When it comes to estimating PA {in general growing networks} without time-resolved data, no satisfactory methods exist. All existing methods assume either unrealistic network types or unnecessarily restrictive functional forms for $A_k$~\cite{timeline_mcmc_icml,sheridan,timeline_adaptive_sampling,wan2017,gao2017,cantwell_2021}. {A method for estimating $A_k$ that does not require time-resolved data stands to advance the field of complex networks on account that there are hundreds of one-snapshot, real-world growing network datasets in online databases waiting to be analyzed.} Such a method would allow researchers to uncover new insights {about} PA and various related phenomena that {may currently lie} in those one-snapshot networks.

Our contributions are two-fold. In our main contribution, {in Section~\ref{sec:result_methods},} we propose a method called PAFit-oneshot to nonparamterically estimate the PA function of a growing network from its final snapshot, $G_T$, alone. Our method does not assume any functional form for the PA function, and can be applied to any real-world network snapshot. At the heart of our solution is a correction for a bias that occurs when the PA function is nonparametrically estimated from one snapshot. Since one always estimates the PA values of the degrees $k$ that exist in the snapshot, the numbers of nodes with those degrees are always positive \emph{a priori}. Failing to account for this bias leads to what may be called a waterfall artefact; that is, a severe underestimation of $A_k$ in the region of large $k$, as we illustrate in Fig.~\ref{fig:illustrative}. {Surprisingly, the presence of this bias, let alone a proposed correction for it, has never been discussed in the literature.}

{To remove this bias, we need to estimate the probability that $k$ exists in $G_T$}. This resembles the problem of selective inference~\cite{Taylor7629,tibshirani2016exactpsi}. {Selective} inference typically considers model selection in a regression setting and adjusts the bias of regression coefficients for the selected predictors; one must correct for the effect of choosing the predictor. In our problem, model selection is equivalent to choosing over which {values} of $k$ {to use to} estimate $A_k$, which is where $k$ exists in $G_T$. {Starting from an initial rough estimation of the probability that we observe $k$ in $G_T$, our method iteratively improves this estimation using Monte Carlo simulations}. The proposed method is implemented in the \textsf{R} package \texttt{PAFit}~\cite{pham_jss}.

In our second contribution, in Section~\ref{sec:real_data_no_timeline}, we applied the proposed method to three real-world networks without growth information, and obtained sublinear PA functions in all three cases. Since these networks do not contain time-resolved data, it had been impossible to estimate PA functions of these networks until now. The sublinear functions are considerably weaker than the conventional linear PA form $A_k = k + c$ which is often employed in modelling one-snapshot networks~\cite{timeline_mcmc_icml,wan2017} or as similarity index in link predictions~\cite{link_prediction_survey}. This emphasizes the need to look beyond the conventional linear PA in one-snapshot networks.

\section{Model and Proposed Method}
\label{sec:theoretical_background}
After introducing the network model {underlying our proposed method} in Section~\ref{sec:model}, we provide {a} theoretical foundation {for the method} in Section~\ref{sec:formula}. We then introduce a baseline method in Section~\ref{sec:baseline} and present our proposed method in Section~\ref{sec:result_methods}. Finally, we demonstrate our method through an example in Section~\ref{sec:illustrative}. {Table~\ref{tab:notation} summarizes the notation used in this paper.}

\begin{table*}[!h]
{
\caption{A reference table of the notation used in this paper. 
{\label{tab:notation}}}
\centering
\begin{tabularx}{\textwidth}{l  X} 
\hline
\textbf{Notation} & \textbf{Meaning} \\ \hline
$k$ & Generic symbol for node degree \\
$t \in \{1, \ldots, T\}$ & Time-step in the evolution of a growing network \\
$T$ & Final time-step in the evolution of a growing network \\
$G_t$ & Growing network at time-step $t$ \\ 
$p(t)$ & Node rate at time-step $t$ \\
$p$ & Limit of the sequence $p(t)$ as $t$ goes to infinity \\
$N(t)$ & Number of nodes in $G_t$\\
$N = N(T)$ & Number of nodes in $G_T$\\
$E(t)$ & Number of edges in $G_t$ \\
$E = E(T)$ & Number of edges in $G_T$ \\
$n_k(t)$ & Number of degree $k$ nodes in $G_t$ \\ 
$\mu_k$ & Limit of the sequence $n_k(t)/(pt)$ as $t$ goes to infinity \\
$A_k$ & Attachment kernel \\
$H(t) = \sum_{k} A_k n_{k}(t)$ & Attachment kernel normalizing constant at time-step $t$ \\ 
$\eta$ & Limit of the sequence $H(t)/t$ as $t$ goes to infinity\\
$p_k = \mathbb{P}(n_{k}(T) > 0)$ & Selection probability of degree $k$ \\
$S$ & Number of simulation rounds used in the PAFit-oneshot method \\
$s \in \{1, \ldots, S\}$ & Round of simulation in the PAFit-oneshot method \\
$M$ & Number of simulated networks in the $s$-th round \\
$\hat{A}_k^{(s)}$ & Estimated value of $A_k$ at the $s$-th round \\
$\hat{p}_k^{(s)}$ & Estimated value of $p_k$ at the $s$-th round \\
$n_k(T)^{(s,i)}$ & Number of nodes with degree $k$ at time-step $T$ in the $i$-th simulated network in the $s$-th round\\
$p_k^{(s,i)}$ & Indicator function of whether $n_k(T)^{(s,i)}$ is positive  \\
$\hat{p}_k$ & Estimated value of $p_k$ after $S$ rounds \\
$\hat{p}^{\text{final}}$ & Final estimated value of $p_k$ \\ 
$\hat{A}_k$ & Final estimated value of $A_k$ \\ \hline
\end{tabularx}
}
\end{table*}

\subsection{The Simple Growth model}\label{sec:model}
The {PAFit-oneshot} method assumes the {novel} Simple Growth (SG) network model. {It} is {a} directed {network model} and starts at time $t = 1$ with two {singleton} nodes. At each time-step~$t>1$, to form $G_{t+1}$, we add to $G_t$ either a new isolated node with probability $p(t)$ or a new edge between existing nodes with probability $1 - p(t)$. In the latter case, the destination node is chosen based on the PA rule: a node with in-degree $k$ is chosen with probability proportional to $A_k$. The quantity $p(t)$ is a number in $(0,1)$. {Although} it is sometimes called an edge-step function~\cite{PA_edge_step}{, in this paper we call it node rate}. 

{The SG model is flexible enough for modelling various} real-world networks. A network generated by the SG model is typically not a tree, since there can be a new edge between existing nodes at each time-step, which can create cycles. The expected {ratio} of the number of nodes {to} edges $\mathbb{E}[N(t)/E(t)]$ in {the SG} model is approximately $\sum_{i=1}^{t-1}p(i)/(t-1-\sum_{i=1}^{t-1}p(i))$, which can be tuned to be any positive value. 

In this paper we are only interested in the PA phenomenon related to the in-degree distribution of a network. Therefore, we do not explicitly model how source nodes are selected and assume only that, conditional on $G_t$, source node selection is independent of how the destination node is chosen. Furthermore, unless stated otherwise, all degree-related quantities should be understood to refer to in-degree. 

\subsection{A formula for estimating $A_k$ from one snapshot}\label{sec:formula}


\bigskip

\noindent
Let $n_k(t)$ be the number of {degree $k$ nodes} at time-step $t$,
{and 
\[
\mu_k = \lim_{t\to\infty} n_k(t)/N(t),
\]
if it exists, be the limit of the degree distribution.
The intuition underlying our approach is that $\{\mu_k\}_{k=0}^\infty$
captures enough information to accurately estimate $\{A_k\}_{k=0}^\infty$. 
The following important lemma, which we prove in Appendix~\ref{sec_sup:proof_theorem_1_rigor}, is obtained from a recursive equation that the $\mathbb{E}n_k(t)$'s must satisfy in the limit of large $t$.}

\begin{lemma}
\label{theorem:mean_converge_in_degree}
Assume that $p(t) = p + \mathcal{O}(t^{-1/2}\log t)$ with $0 < p < 1$ and $\{A_k\}_{k=0}^\infty$ satisfies the conditions stipulated below. For all $k\ge 0$, let {the} $\mu_k$'s be the constants {satisfying}
\begin{equation}
  \mu_k = \lambda^{-1}(A_{k-1}\mu_{k-1} - A_k\mu_k) + \boldsymbol{1}_{k = 0},\label{eq:mu_recursive}  
\end{equation}
with {$\lambda = \eta(1-p)^{-1}$ for the constant $\eta$ in assumption (b) below and with the convention} that $A_{-1}=\mu_{-1}=0$. For a fixed $k$, there exist constants {$m^{*}$ and $t^{*}$} such that for all $j \le k$, $\lvert \mathbb{E}[n_{j}(t)/(pt)] - \mu_j \rvert \le m^{*} t^{-1/2}\log t$ for all $t\ge t^{*}$, {meaning that $\{\mu_k\}_{k=0}^\infty$ is the limit of the degree distribution.
Equation~(\ref{eq:mu_recursive}) can be used to check that $\mu_k$'s form a probability distribution, i.e.  $\mu_k > 0$ and $\sum_k \mu_k = 1$.}
\end{lemma}

For Lemma~\ref{theorem:mean_converge_in_degree}, we assume that
\begin{enumerate}[label=(\alph*)]
\item $A_k \le A_{k+1}$ for all $k$ and, 
\item 
There exists a constant $\eta > 0$ such that $\mathbb{P}(\lvert H(t)/t - \eta \rvert \ge t^{-1/2}\log t) \le \mathcal{O}(t^{-1/2}\log t)$, with $H(t) = \sum_{k} A_k n_{k}(t)$. 
\end{enumerate}

The first assumption {(a)} means that $A_k$ is a non-decreasing function, which is true for the power-law form $A_k = k^\alpha$ with $\alpha \ge 0$ or the linear form $A_k = k + c$ with $c > 0$. This assumption has also been used in previous PA network models~\cite{non_decreasing_PA_example,dereich2009}. While we need this assumption for our convergence proof, PAFit-oneshot does not actually employ this assumption in its procedure and, based on our experience, works even when the true PA function is not monotone. 

The second assumption {(b) implies that $H(t)/t$ converges to some constant $\eta$. We note that the actual value of $\eta$ is not needed for estimating $A_k$.} This assumption is needed in order to handle the normalizing factor $H(t)$. A similar version of this assumption has been used previously~\cite{krapi2}. The assumption is satisfied, for example, by $A_k = k + c$ with $\eta = 1 + (c-1)p$ or $A_k = 1$ with $\eta = 1 - p$. Note that in these two cases, $\mathbb{P}(\lvert H(t)/t - \eta \rvert \ge t^{-1/2}\log t) \le \mathcal{O}(t^{-b})$ for any positive constant $b$, which is a much stronger rate than what is assumed in the second assumption. {While we suspect that this assumption holds for all sub-linear and linear PA functions, the rigorous investigation is left to future work.}

Equation~(\ref{eq:mu_recursive}) is similar to {an} equation in the case of PA trees~\cite{krapi,rudas_2007}. {In such models, the constant $\lambda$ is often called the Malthusian parameter due to a connection with continuous-time branching processes.}


Using Lemma~\ref{theorem:mean_converge_in_degree}, we {prove} the following theorem which is key to the derivation of the PAFit-oneshot method. Gao et al.~\cite{gao2017} provided a similar formula for PA trees.
\begin{theorem}\label{theorem:mean_converge}
{Assume the same conditions as in Lemma~\ref{theorem:mean_converge_in_degree}. F}or a fixed $k$ we have
\begin{equation}
A_k = \lambda\dfrac{\mathbb{E}\sum_{j > k}n_{j}(t)}{\mathbb{E}{n_{k}}(t)} + o(1), \label{eq:si_basic_equation}
\end{equation}
as $t$ tends to infinity, where $\lambda$ is some constant that is independent of $t$ and $k$.
\end{theorem}

\begin{proof}
The following equation can be {derived} from Eq.~(\ref{eq:mu_recursive}):
\begin{equation}
  \sum_{j > k}\mu_j = \lambda^{-1}A_k\mu_{k},\label{eq:mu_recursive_2}  
\end{equation}
for $k \ge 0$.
We then have:
\begin{align}
\mathbb{E}\sum_{j  > k + 1}n_j(t) &= \mathbb{E}\left[N(t) - \sum_{j=0}^{k}n_j(t)\right] = \sum_{i = 1}^{t-1}p(i) + 2 - \sum_{j=0}^{k}\mathbb{E}n_j(t)\nonumber\\
&= (t-1)p + 2 - tp\sum_{j = 0}^{k}\mu_k + \mathcal{O}(t^{1/2}\log t) \quad \text{(using Lemma~\ref{theorem:mean_converge_in_degree})}\nonumber\\
&= tp\sum_{j > k}\mu_j + \mathcal{O}(t^{1/2}\log t) \quad \text{ (using $\sum_j \mu_j = 1$)}\nonumber\\
&= tp\dfrac{A_k\mu_k}{\lambda} + \mathcal{O}(t^{1/2}\log t) \quad \text{ (using Eq.~(\ref{eq:mu_recursive_2}))}\nonumber\\
&= \dfrac{A_k\mathbb{E}n_{k}(t)}{\lambda} + \mathcal{O}(t^{1/2}\log t) \quad \text{(using Lemma~\ref{theorem:mean_converge_in_degree})}.\nonumber
\end{align}
Noting that, from Lemma~\ref{theorem:mean_converge_in_degree}, $\mathbb{E}n_k(t)$ is positive for sufficiently large $t$, we have:
\begin{align}
\dfrac{\mathbb{E}\sum_{j > k}n_j(t)}{\mathbb{E}n_{k}(t)} &=  \dfrac{A_k}{\lambda} + \dfrac{\mathcal{O}(t^{1/2}\log t)}{\mathbb{E}n_{k}(t)} =  \dfrac{A_k}{\lambda} + \mathcal{O}(t^{-1/2}\log t).
\end{align}
This concludes the proof of Theorem~\ref{theorem:mean_converge}.
\end{proof}

\subsection{A baseline method}
\label{sec:baseline}
{Here} we derive an estimation method for $A_k$ from Eq.~(\ref{eq:si_basic_equation}). Assuming that $T$ is large enough, we obtain:
\begin{equation}
A_k \approx \lambda\dfrac{\mathbb{E}\sum_{j > k}n_{j}(T)}{\mathbb{E}{n_{k}}(T)}, \label{eq:si_basic_equation_2}
\end{equation}
with $\lambda$ being a constant that is independent of $k$ and hence {may be safely} ignored. One then estimates $\mathbb{E}\sum_{j > k}n_{j}(T)$ and $\mathbb{E}{n_{k}}(T)$ by $\sum_{j > k}n_{j}(T)$ and $n_{k}(T)$, respectively. This yields the baseline method:
\begin{equation}
\hat{A}_k^{\text{baseline}} = \dfrac{\sum_{j >k}n_{j}(T)}{n_{k}(T)}. \label{eq:baseline}
\end{equation}
While this estimator works very well for small $k$, for large $k$ it suffers from {a} waterfall artefact: the estimated values of $A_k$ fall off rapidly ({see} Fig.~\ref{fig:illustrative}). To our knowledge, while this estimator has been proposed {only} for PA trees by Gao et al.~\cite{gao2017}, it has never been applied to any real-world networks, and the waterfall artefact, let alone its cause, has never been discussed in the literature. 

\subsection{{PAFit-oneshot: A n}ew method for estimating $A_k$ from one snapshot}
\label{sec:result_methods}
We present our main contribution: the novel PAFit-oneshot method for estimating~$A_k$ when only $G_T$, the snapshot {network} at time-step $T$, is observed. At the heart of our method is a correction to the waterfall artefact inherent to Eq.~(\ref{eq:baseline}).

The root of the waterfall artefact is that $n_k(T)$ is {a poor} estimator of $\mathbb{E}{n_{k}}(T)$ when~$k$ is large. Given $G_T$, we only estimate the $A_k$ values for the degrees $k$ observed in $G_T$, which means $n_{k}(T)$ is positive \emph{a priori}. Therefore, $n_{k}(T)$ is actually an estimator for the conditional expectation $\mathbb{E}\left[n_{k}(T) \mid n_{k}(T) > 0\right]$, which is equal to $\mathbb{E}n_{k}(T) / \mathbb{P}(n_{k}(T) > 0)$. {Let $p_k$ denote the selection probability $\mathbb{P}(n_{k}(T) > 0)$.} Correcting for $p_k$ leads us to the following equation:
\begin{equation}
A_k \approx \dfrac{\sum_{j > k}n_{j}(T)}{n_{k}(T)p_k}. \label{eq:mcPA-SI-true}
\end{equation}
Since $A_k$ is only identifiable up to a multiplicative constant, were the $p_k$ values the same for all $k$, Eq.~(\ref{eq:mcPA-SI-true}) would be equivalent to Eq.~(\ref{eq:si_basic_equation_2}). However, since $p_k$ tends to decrease rapidly when $k$ is large ({see} Fig.~\ref{fig:illustrative}{(b)}), the correction in Eq.~(\ref{eq:mcPA-SI-true}) is necessary.

Correcting for the probability of observing $k$ resembles the problem of post-selection inference~\cite{Taylor7629,tibshirani2016exactpsi}. Post-selection inference adjusts for {a} bias that crops up from the act of parameter selection. One {typically} achieves the bias-correction by changing all probabilities in the subsequent data analysis to conditional probabilities that are conditioned on the selection event. In {the present setting where we are} estimating~$A_k$, we consider the conditional probabilities given that the degree $k$ is observed in $G_T$, i.e., $n_k(T) > 0$.

We estimate {the selection probability} $p_k$ via $S \ge 1$ rounds of simulations for sequentially updating $\hat{A}_k^{(s)}$ and $\hat{p}_k^{(s)}$, $s=1,\ldots,S$.
We start with the same baseline estimator {as} in Eq.~(\ref{eq:baseline}), which is equivalent to initially estimating $p_k \propto 1$ in Eq.~(\ref{eq:mcPA-SI-true}). In particular, we define~$\hat{A}_k^{(0)}$, the initial estimate of our iterative procedure, to be equal to $\hat{A}_k^{\text{baseline}}$. At each round $s = 1,\ldots,S$, we use $\hat{A}_k^{(s-1)}$ as the true PA function and simulate $M$ networks. {Let $n_k(T)^{(s,i)}$ denote the number of nodes with degree $k$ at time-step $T$ in the $i$-th simulated network of the $s$-th round.} We see whether the degree $k$ exists at time-step $T$ in the $i$-th simulated network:
\begin{equation}
p_{k}^{(s,i)} \coloneqq \mathbf{1}_{n_{k}(T)^{(s,i)} > 0},\label{eq:p_k_each_sim}
\end{equation}
and estimate $\hat{p}_k^{(s)}$ by {averaging} the $p_k^{(s,i)}$'s:
\begin{equation}
\hat{p}_{k}^{(s)} \coloneqq \dfrac{1}{M}\sum_{i = 1}^{M}p_{k}^{(s,i)}.\label{eq:p_k_round_s}
\end{equation}
When simulating SG model networks, we assume that $p(t) = \hat{p}$ with $\hat{p} = (N-2)/(E + N - 2)$, where $N$ and $E$ are the numbers of nodes and edges in the observed network $G_T$, respectively. At the end of round $s$, we update $\hat{A}_k^{(s)}$ using Eq.~(\ref{eq:mcPA-SI-true}) with the current value of $\hat{p}_k^{(s)}$:
\begin{equation}
\hat{A}_k^{(s)} = \dfrac{\sum_{j > k}n_{k}(T)}{n_k(T)\hat{p}_{k}^{(s)}}. \label{eq:mcPA_SI_round_s}
\end{equation}
After $S$ rounds, the estimate for $p_k$ is the average of the $\hat p_{k}^{(s)}$'s:
\begin{equation}
\hat{p}_{k} \coloneqq \dfrac{1}{S}\sum_{s = 1}^{S} \hat{p}_{k}^{(s)}.\label{eq:final_p_k}
\end{equation}
For additional stability, we repeat the whole process five times to obtain five values of $\hat{p}_k$ and use their average, denoted as $\hat{p}_k^{\text{final}}$, as the final estimate of $p_k$.
The final estimate of $A_k$ is:
\begin{equation}
\hat{A}_k = \dfrac{\sum_{j > k}n_{k}(T)}{n_k(T)\hat{p}_{k}^{\text{final}}}. \label{eq:mcPA_SI_final}
\end{equation}

The number of simulated networks, $M$, in each round and the number of simulation rounds, $S$, are parameters in PAFit-oneshot. In our experience, while  an $M$ of around $100$ is enough, a small value of $S$, for example, $S = 5$, is stable.

A schematic presentation of PAFit-oneshot is given in Fig.~\ref{fig:mcPA-SI-fig}. In order to avoid clutter, we leave the binning version of the method to Appendix~\ref{sup_sec:binning}.
{
\subsection{Related work}
There is a large body of literature on the theory of PA-based network models. Some of the first rigorous treatments of some general models are given in~\cite{bollobas_2001,bollobas_directed_graph,cooper_model}. Some notable models that allow a stochastic number of new edges at each time-step are~\cite{cooper_model,sheridan_2008,deijfen2009,feng_2017,feng_2020,PA_edge_step}. The degree distribution for the case of a general $A_k$ has been derived for various models~\cite{krapi,rudas_2007,dereich2009}. Equations that are similar to Eqs.~(\ref{eq:si_basic_equation}) and~(\ref{eq:mu_recursive_2}) are given in the works of Gao et al.~\cite{gao_2016,gao2017}.}
{
There is also a considerable amount of studies on estimating the PA function from empirical data. When time-resolved data is available, a large number of methods have been proposed~\cite{jeong,newman2001clustering, massen,Gomez,pham2}. When there is, however, only one snapshot, to our knowledge, there is no existing method capable of nonparametrically estimating the PA function for an arbitrary network. All previous works assume either a restrictive network model, e.g., trees~\cite{sheridan,gao2017,cantwell_2021}, or some simplistic PA model, namely $A_k = k + c$~\cite{timeline_mcmc_icml,timeline_adaptive_sampling,gao_2016,wan2017}. The nonparametric method of Gao et al.~\cite{gao2017}, which was proposed only for PA trees, gives an estimation formula that is the same as the baseline method. This method, however, suffers from the aforementioned waterfall artifact when applied to trees.}

\section{Simulation Study}
\label{sec:sim_study}
{We start the section by demonstrating our method with an illustrative example in Section~\ref{sec:illustrative}. We then perform a systematic investigation of the performance of PAFit-oneshot} in two types of simulations: in Section~\ref{sec:sim_constant_p}, {the node rate} $p(t)$ is held fixed, while in Section~\ref{sec:sim_varrying_p} {it} varies based on real-world data.

To generate networks, we use the power-law yielding form $A_k = \max (k,1)^{\alpha}$, which has been used frequently in previous works. {For Section~\ref{sec:sim_constant_p} and~\ref{sec:sim_varrying_p}, w}e investigate five values of {attachment exponent} $\alpha$: 0, 0.25, 0.5, 0.75, and 1. From the {estimated} value $\hat{A}_k$ of each method, we estimate $\alpha$ and use the quality of the result as a proxy to judge how well each method estimates $A_k$. To estimate $\alpha$ from $\hat{A}_k$, we employ least squares for the baseline method, and a weighted least squares method for PAFit-oneshot, since this method provides standard deviations for the individual $\hat{A}_k$ values{; the weight is set to inversely proportional to the variance of $\hat{A}_k$.}

An advantage of our nonparametric approach is that we do not need to pre-specify a functional form for $A_k$; one can simply fit a specific form to the estimated value $\hat{A}_k$. Beside the power-law form, in Appendix~\ref{sec_sup:sim} we use another functional form as the true PA function.

\subsection{An illustrative example}
\label{sec:illustrative}
\begin{figure*}[!h]
  \centering
  \includegraphics[width=\textwidth]{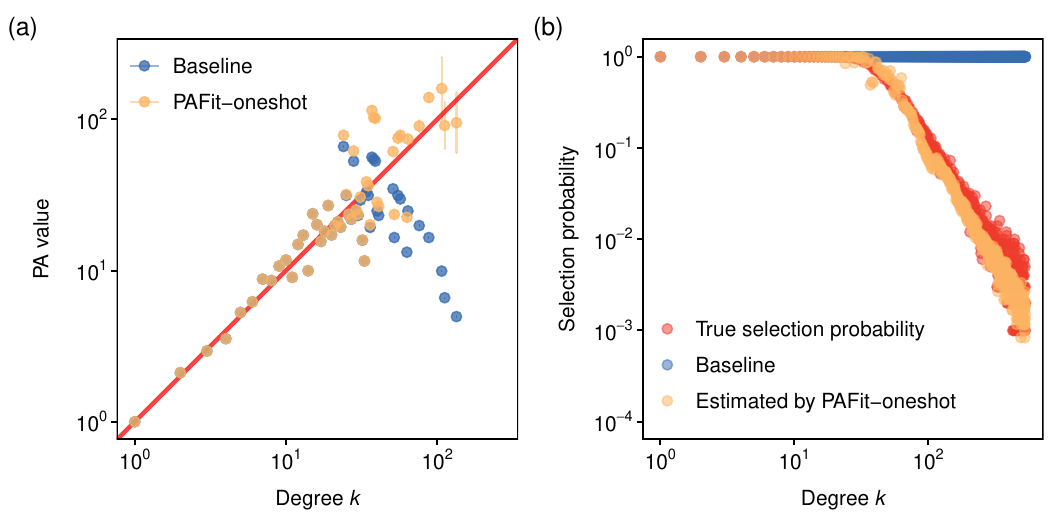}
  \caption{Estimating $A_k$ from one network snapshot of an SG model generated network with true PA function is $A_k = \max(k,1)^\alpha$ with $\alpha = 1$. {(a)} The proposed method (yellow dots) corrects the waterfall artefact in the baseline method estimate (blue dots). The bar at each yellow dot depicts the plus/minus two-sigma confidence interval estimated by the proposed method. The true PA function is shown as a red line for reference. We obtained $\hat{\alpha}_{\text{baseline}} =  0.52~(\pm 0.19)$ and $\hat{\alpha}_{\text{PAFit-oneshot}}  = 0.93~(\pm 0.08)$; the confidence interval for $\hat{\alpha}$ is plus/minus two-sigma. {(b)} PAFit-oneshot improves upon the baseline in estimating {the selection probabilities} $p_k =\mathbb{P}(n_k(T) > 0)$. The true $p_k$ is calculated from $1000$ simulated SG model networks. For ease of visualization, all the series are scaled so that the value of each series at $k = 1$ is $1$. \label{fig:illustrative}}
\end{figure*}

Figure~\ref{fig:illustrative} showcases our method through a simple simulated example that uses the linear PA function $A_k = \max{(k,1)}$ as the true PA function. The underlying network was generated by the SG model with $p(t) = 0.5$ and $T = 50000$. We applied PAFit-oneshot with $M = 200$ and $S = 5$.

Figure~\ref{fig:illustrative}{(a)} shows that PAFit-oneshot accurately recovers the true PA function. By contrast, the baseline estimate given by Eq.~(\ref{eq:baseline}) performs well when $k$ is small, but is plagued by the waterfall artefact when $k$ is large. To our knowledge, no principled way to automatically determine where the waterfall artefact starts to kick in has yet been devised. Due to this artefact, estimating $\alpha$ using the least squares method from the entire range of the estimation result $\hat{A}_k$ leads to a severely underestimated value: $\hat{\alpha}_{\text{baseline}} = 0.52~(\pm 0.19)${, while the true value is $\alpha=1$.}

This artefact occurs since the baseline method does not take into account the \emph{a pirori} existence of the degree $k$ nodes in the snapshot. The cause of the waterfall artefact is that the baseline method {assumes $p_k=1$ which} over-estimates $p_k$ for large $k$. Using a simulation step to estimate $p_k$ leads to better estimation of $p_k$, as can be seen in Fig.~\ref{fig:illustrative}{(b)}. Using a weighted least squares method, we obtain $\hat{\alpha}_{\text{PAFit-oneshot}}  = 0.93~(\pm 0.08)${, which is a good estimate of $\alpha=1$.}

\subsection{Constant {node rate} $p(t)$}
\label{sec:sim_constant_p}
\begin{figure*}[!h]
  \centering
  \includegraphics[width= \textwidth]{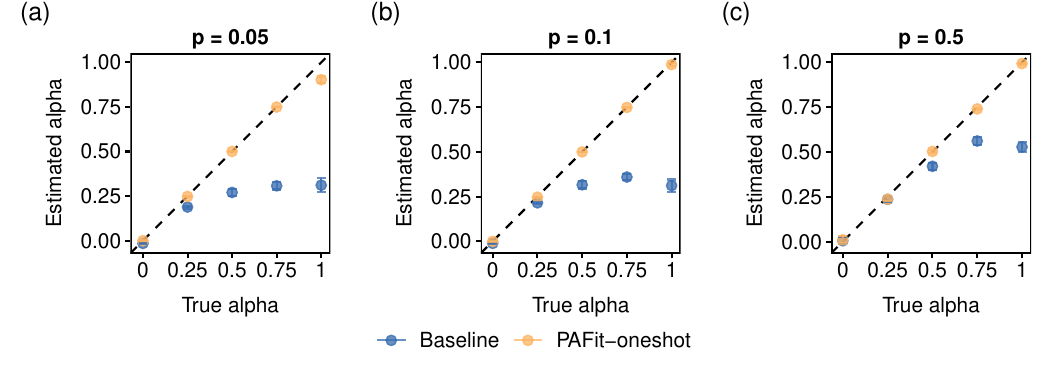}
  \caption{Estimated $\alpha$ from $\hat{A}_k$. For each case, $p(t) = p$, i.e., a constant. The network has $T = 5\times 10^5$ and $A_k = \max (k,1)^{\alpha}$ as the true PA function. In each combination of the parameters $p$ and $\alpha$, for each method we plot the mean of the estimated $\alpha$ and its plus/minus two-sigma confidence interval obtained from $50$ simulated networks. {(a) $p = 0.05$. (b) $p = 0.1$. (c) $p = 0.5$.}}\label{fig:simulation}
\end{figure*}

In this section, we assume $p(t) = p$ and test three values of $p$: $p =0.05$, $p =0.1$, and $p =0.5$. These values are similar to the average values of $p(t)$ observed in real-world networks (see Tables~\ref{sup_tab: dataset_full_timeseries} and~\ref{sup_tab: dataset_no_timeseries_stats}). A total of $50$ networks were simulated for each value of $p$. The total number of time-steps in each network is $T = 5\times 10^5$. The estimation result for each $\alpha$ is shown in Fig.~\ref{fig:simulation}.

The proposed method outperforms the baseline method in all settings, since the baseline method severely underestimate $\alpha$ due to the waterfall artefact. PAFit-oneshot underestimates $\alpha$ when $\alpha$ is near 1 and $p = 0.05$. When $p$ is small, there will be more new edges and less new nodes added to the network. This will often increase the maximum degree in the network when $\alpha$ is near $1$, which makes the estimation harder since there are more $A_k$ values to be estimated. Interestingly, the same phenomenon has been observed in estimating the PA function with full time information~\cite{pham2,pham3,pham_jss}. Taking also the results in Fig.~\ref{fig:simulation_redner} into consideration, one can then conclude that the proposed method works well and outperforms the baseline when $p(t)$ is constant.

\subsection{Varying {node rate} $p(t)$}
\label{sec:sim_varrying_p}
Here we check by simulations the robustness of the proposed method, which assumes $p(t) = p$ for the simulation step, in real-world situations when $p(t)$ varies over time. In {each} simulation, at each time-step, we added a new node or a new edge according to a node-edge sequence taken from a real-world network, and only sampled where the new edge connects based on the PA rule $A_k = \max (k,1)^\alpha$. We used three networks: the Enron email network~\cite{enron}, the Escort rating network~\cite{escort}, and the UCIrvine forum message network~\cite{ucirvine}. Some of their statistics {are} found in Table~\ref{sup_tab: dataset_full_timeseries}. 

In all three networks, the estimated $p(t)$ function varies greatly, as can be seen from Fig.~\ref{sup_fig:real_p}. While the $p(t)$ {sequence} in the Escort network and the UCIrvine network start high then rapidly decrease to some stable value around $0.1$, the series of Enron networks, while decreasing from a high starting value, show great fluctuations even toward the end of the growth process.

\begin{figure*}[!h]
  \centering
  \includegraphics[width= \textwidth]{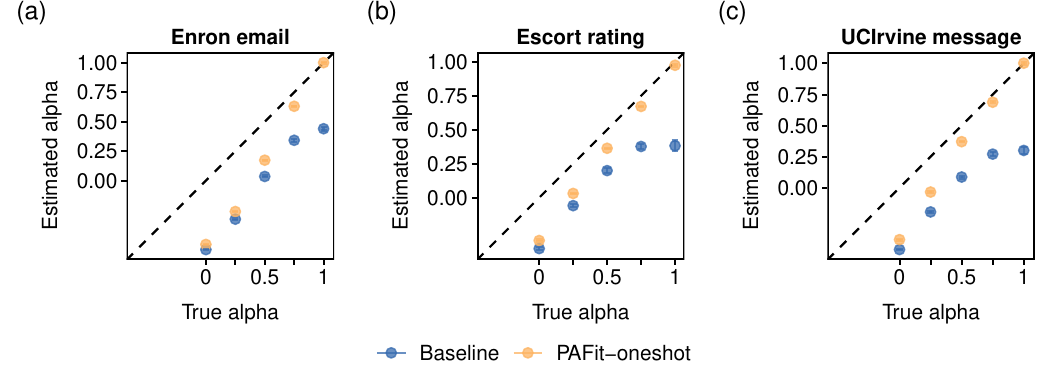}
  \caption{Estimated $\alpha$ from $\hat{A}_k$. We used node-edge sequences obtained from real-world networks and $A_k = \max (k,1)^{\alpha}$. In each setting, for each method we plot the mean of the estimated $\alpha$ and its plus/minus two-sigma confidence interval obtained from $50$ simulated networks. {(a) Enron email. (b) Escort rating. (c) UCIrvine message.}}\label{fig:simulation_p_t_real_log_linear}
\end{figure*}

The results of {these simulations} are shown in Fig.~\ref{fig:simulation_p_t_real_log_linear}. The proposed method again outperforms the baseline. Although {we observe some} performance degradation as expected, PAFit-oneshot performs reasonably well in the region of moderate to large PA effect ($0.75 \le \alpha \le 1$) in all three networks. Figures~\ref{fig:one_example_enron}{(a)}~and~{(b)} show two runs which are typical for the situations when $\alpha$ is small and when $\alpha$ is large, respectively, in the Enron network. When $\alpha$ is small, PAFit-oneshot fails to remove the waterfall artefact and behaves similarly to the baseline method. When $\alpha$ is large, the proposed method successfully removes the waterfall artefact and estimate the PA function comparatively well. Taking also the results in Fig.~\ref{fig:simulation_p_t_real_redner} into account, we conclude that when $p(t)$ is varying, the proposed method outperformed the baseline method in all settings and is reasonably good in the region of moderate to large PA effect.

\section{{PAFit-oneshot recovers the true PA functions in time-resolved real-world networks}}
\label{sec:real_data_with_timeline}
\begin{figure*}[!h]
  \centering
  \includegraphics[width= \textwidth]{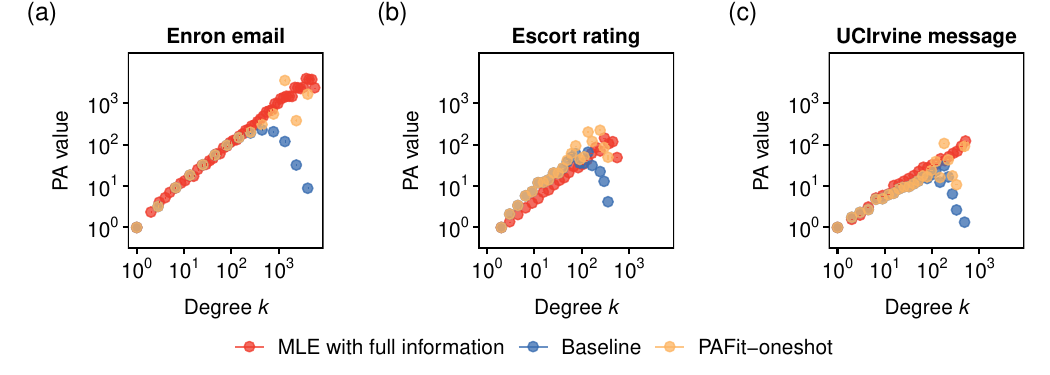}
  \caption{Estimation results in real-world networks when the time-step information is available. {(a) Enron email. (b) Escort rating. (c) UCIrvine message.}}\label{fig:validating_real_PA}
\end{figure*}

We compare the proposed method with the Maximum Likelihood Estimation (MLE), explained in Appendix~\ref{sec:mle_full}. The MLE can be viewed as the gold standard approach when fully time-resolved data is available. We used the three real-world networks we introduced in Section~\ref{sec:sim_varrying_p}: the Enron email, the Escort rating, and the UCIrvine forum message network. For each network, we apply PAFit-oneshot with $M = 50$ and $S = 5$. The results are shown in Figure~\ref{fig:validating_real_PA}.

Compared to the MLE with full information, PAFit-oneshot and the baseline estimated $A_k$ reasonably well when $k$ is small. When $k$ is large, the baseline, however, severely underestimated~$A_k$ as expected. The proposed method did not suffer from the bias and estimated~$A_k$ reasonably well even when~$k$ is large.

Since all the estimated PA functions in three networks {are approximately linear on a log-log scale}, we fitted the functional form $A_k = \max(k,1)^\alpha$ to the estimated $A_k$ and estimated the attachment exponent $\alpha$ ({see} Table~\ref{tab:real_has_timeline}) by weighted least squares. Overall, the proposed method agrees reasonably well with the MLE with full time information on the estimated value of $\alpha$.

\begin{table*}[!h]
\caption{Estimated $\alpha$ in real-world networks with time-resolved data. Each confidence interval of the estimated $\alpha$ is plus/minus two-sigma. The functional form $A_k = \max (k,1)^\alpha$ is fitted to $\hat{A}_k$ by weighted least squares. 
{\label{tab:real_has_timeline}}}
\centering
\begin{tabularx}{\textwidth}{l  l l X}
\hline
\textbf{Estimation method} & \textbf{Enron email} & \textbf{Escort rating} & \textbf{UCIrvine message} \\ \hline 
MLE with {time-resolved data} & $0.93 \pm 0.02$ & $0.92 \pm 0.03$ & $0.74 \pm 0.03$\\ 
PAFit-oneshot & $1.01 \pm 0.05$ & $1.07 \pm 0.09$ & $0.67 \pm 0.04$ \\
Baseline & $0.42 \pm 0.30$ & $0.57 \pm 0.30$ & $0.28 \pm 0.20$ \\ \hline
\end{tabularx}
\end{table*}

\section{{PAFit-oneshot predicts sublinear PA functions in real-world networks with unknown timelines}}
\label{sec:real_data_no_timeline}
\begin{figure*}[!h]
  \centering
  \includegraphics[width= \textwidth]{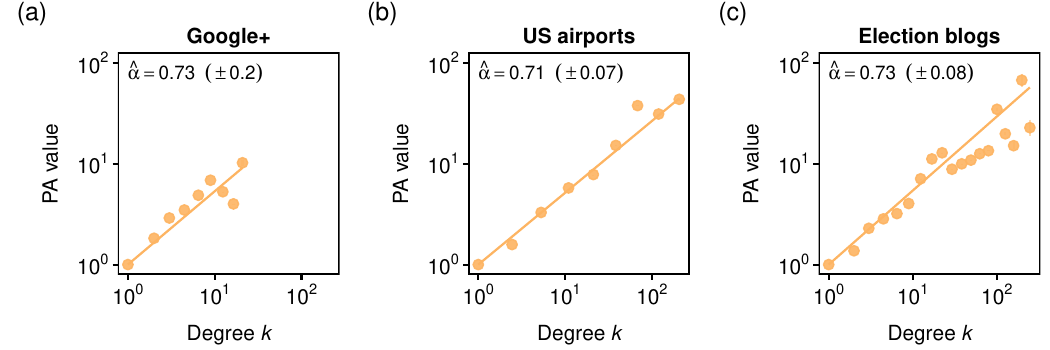}
  \caption{Estimation results in real-world networks with no time-resolved data. The functional form $A_k = \max (k,1)^\alpha$ is fitted to $\hat{A}_k$ by weighted least squares. The confidence interval for each estimated $\alpha$ is plus/minus two-sigma. {(a) Google+. (b) US airports. (c) Election blogs.}}\label{fig:unmeasurable_real_PA}
\end{figure*}

We apply PAFit-oneshot to three networks for which no time-resolved data is presently available: a portion of the Google+ user-user network~\cite{gplus}, a network of {United States airports}~\cite{tore_usairport}, and a followship network between political blogs in the 2004 {United States presidential election}~\cite{us_blog}. For more information on these networks, see Appendix~\ref{sec_sup:unmeasurable}. Since these networks are single-snapshots without time-resolved data, nonparametrically estimating the PA function from them has been impossible up to now. In each network, we apply PAFit-oneshot with $M = 50$ and $S = 5$. The results are shown in Fig.~\ref{fig:unmeasurable_real_PA}. 

 The PA phenomenon is present in all networks. Furthermore, since each estimated $A_k$ is roughly linear on a log-log scale, we fit the functional form $A_k = \max (k,1)^\alpha$ and estimate $\alpha$. Since all estimated $\alpha$ is smaller than $1$, the PA phenomenon in all three networks is sublinear. The estimated PA functions are significantly weaker than the conventional linear form $A_k = k + c$ which is often employed in modelling one-snapshot networks~\cite{timeline_mcmc_icml,wan2017} or as similarity index in link predictions~\cite{link_prediction_survey}. This emphasizes the need to look beyond linear PA in one-snapshot networks. 
 
\section{Conclusion}\label{sec:conclusion}
We proposed a novel method, called PAFit-oneshot, for estimating the PA function nonparametrically from a single network snapshot. The key to our method is a correction of a previously unnoticed bias that has a connection with post-selection inference. We demonstrated that PAFit-oneshot recovers the PA function reasonably well under realistic settings. By applying PAFit-oneshot to three real-world networks without time-resolved data, we found the first evidence for the presence of sublinear PA in real-world, {one-snapshot} network data.  

{There are many directions for improving the current methodology. Firstly, while this work needs assumption (b) of Lemma~\ref{theorem:mean_converge_in_degree} for theoretical analysis, we were able to show it only for the form $A_k = k + c$. Therefore, it is interesting to see this assumption shown for general classes of $A_k$ functions in general growing networks. Secondly, when the node rate $p(t)$ is a constant, the simulation step is demonstrated empirically to work well. It is then important to back this up with a theoretical analysis in future work. One potential starting point is to investigate the selection probability $\mathbb{P}(n_k(T) > 0)$. Thirdly, when the node rate $p(t)$ varies, the method is able to remove the waterfall artefact only when the PA effect is strong. This weakness potentially stems from the fact that we use a constant $p(t)=p$ sequence in the simulation step. While it may be impossible to remove the artefact in this case, it may be feasible to derive a threshold from where $\hat{A}_k^{\text{baseline}}$ starts to deteriorate. Lastly, the proposed method may be extended to problems of estimating other network growth mechanisms by investigating the presence of selective biases in these problems.}

On the application front, our work opens up a path to estimate the PA function in previously out-of-reach one-snapshot networks. It is our hope that the incoming new evidence provides new insights into the connection between the processes underlying complex network evolution and their large-scale topological features. 

\section*{Funding}
This work was supported by the Japan Society for the Promotion of Science KAKENHI [JP19K20231 to T.P., JP20H04148 to H.S.].


\appendix
\section{Proof of Lemma~\ref{theorem:mean_converge_in_degree}}
\label{sec_sup:proof_theorem_1_rigor}

Recall that $H(t) = \sum_{k}n_k(t)A_k$ is the normalizing factor at time-step $t$. We have:
\begin{align}
\mathbb{E}[n_k(t+1)| G_t] &= n_k(t) + (1-p(t))A_{k-1}n_{k-1}(t)/H(t) - (1-p(t))A_kn_k(t)/H(t) \nonumber \\&+ p(t)\boldsymbol{1}_{k = 0}.\label{theorem_1:cond_mean}
\end{align}
The quantity $A_{k}n_{k}(t)/H(t)$ is $\mathcal{O}(1)$ for all $k \ge 0$ and $t \ge 1$. The error of approximating $A_{k}n_{k}(t)/H(t)$ by $A_{k}n_{k}(t)/(\eta t)$ is $\Delta_k = \left((\eta t - H(t)) A_{k}n_{k}(t)/H(t)\right)/(\eta t)$, which is always $\mathcal{O}(1)$.  Due to assumption (b), $\mathbb{E}\Delta_k = \mathcal{O}(t^{-1/2}\log t)$. The same holds for $\mathbb{E}\Delta_{k-1}$. 

Taking expectations of both sides of Eq.~(\ref{theorem_1:cond_mean}), we then have:
\begin{align}
\mathbb{E}n_k(t+1) &= \mathbb{E}n_k(t) + (1-p(t))A_{k-1}\mathbb{E}n_{k-1}(t)/(\eta t) - (1-p(t))A_k\mathbb{E}n_k(t)/(\eta t)\nonumber \\  &+ p(t)\boldsymbol{1}_{k = 0} +\mathcal{O}(t^{-1/2}\log t).
\end{align}
Substituting $p(t) = p + \mathcal{O}(t^{-1/2}\log t)$ into the preceding equation yields:
\begin{align}
\mathbb{E}n_k(t+1) &= \mathbb{E}n_k(t) + \dfrac{A_{k-1}}{\lambda t}\mathbb{E}n_{k-1}(t) - \dfrac{A_k}{\lambda t}\mathbb{E}n_k(t) + p\boldsymbol{1}_{k = 0} + \mathcal{O}(t^{-1/2}\log t)
, \label{theorem_1:mean}
\end{align}
for all $k\ge 0$. 

One can show heuristically that Eq.~(\ref{theorem_1:mean}) implies Eq.~(\ref{eq:mu_recursive}) by dividing both sides of Eq.~(\ref{theorem_1:mean}) by $p$ and re-arranging the terms:
\begin{align}
(t+1)\dfrac{\mathbb{E}n_k(t+1)}{p(t+1)}  - t\dfrac{\mathbb{E}n_k(t)}{pt} &= \lambda^{-1}\left(A_{k-1}\dfrac{\mathbb{E}n_{k-1}(t)}{pt} - A_k\dfrac{\mathbb{E}n_k(t)}{pt}\right) + \boldsymbol{1}_{k = 0} \nonumber \\  &+ \mathcal{O}(t^{-1/2}\log t).\label{theorem_1:mean_temp}
\end{align}
If one assumes that $\lim\limits_{t\rightarrow \infty}\mathbb{E}n_k(t)/(pt)$ exists and is equal to $\mu_k$, sending $t$ to infinity in Eq.~(\ref{theorem_1:mean_temp}) yields Eq.~(\ref{eq:mu_recursive}). 

While the {following} rigorous argument is more involved, it is standard. Let $\epsilon_k(t) = \mathbb{E}n_k(t) - pt\mu_k$. From Eq.~(\ref{eq:mu_recursive}):
\begin{align}
(t+1)p\mu_k &= tp\mu_k + p\mu_k = tp\mu_k + p\lambda^{-1}A_{k-1}\mu_{k-1} -p\lambda^{-1}A_k\mu_k + p1_{k = 0} \nonumber \\
&= tp\mu_k +\dfrac{A_{k-1}}{\lambda t}pt\mu_{k-1} -\dfrac{A_k}{\lambda t}pt\mu_k + p1_{k=0}.
\end{align}
Subtracting this from Eq.~(\ref{theorem_1:mean}) gives us:
\begin{equation}
\epsilon_k(t+1) = \epsilon_k(t) + \dfrac{A_{k-1}}{\lambda t}\epsilon_{k-1}(t) - \dfrac{A_{k}}{\lambda t}\epsilon_k(t) + \mathcal{O}(t^{-1/2}\log t),\label{theorem_1:mean_3}
\end{equation}
for all $k \ge 0$. This means that there exist some constants $L \ge 0$ and $t_0 \ge 1$ such that:
\begin{equation}
\left\lvert \epsilon_k(t+1) - \left(1 - \dfrac{A_{k}}{\lambda t}\right)\epsilon_k(t) - \dfrac{A_{k-1}}{\lambda t}\epsilon_{k-1}(t) \right\rvert \le Lt^{-1/2}\log t,\label{theorem_1:mean_inequality}
\end{equation}
for all $t \ge t_0$ and all $k \ge 0$.

For a fixed $k$, we will prove that there exist constants {$m^{*}$ and $t^{*}$}, to be specified later,  such that 
\begin{equation}\lvert \epsilon_j(t) \rvert \le m^{*}\sqrt{t}\log t,
\end{equation}
for $t\ge t^{*}$ and $0 \le j \le k$ by induction on $t$. This will prove the assertion.

When $t = t^{*}$, for all $j$, we have $\epsilon_j(t^{*}) \le \mathbb{E}n_j(t^{*}) + pt^{*}\mu_j \le t^{*} + 1 + t^{*}$, thus $\epsilon_j(t^{*}) \le 2t^{*} + 1$, which is at most $m^{*}\sqrt{t^{*}}\log t^{*}$ if we choose $m^{*} \ge \left((2t^{*} + 1)/\sqrt{t^{*}}\right)/\log t^{*}$.

Assume the induction hypothesis for $t \ge t^{*}$, which means that $\lvert \epsilon_j(t) \rvert \le m^{*}\sqrt{t}\log t$ for all $j \le k$ . We will prove that $\lvert \epsilon_j(t + 1) \rvert \le m^{*}\sqrt{t+1}\log (t+1)$ for all $j \le k$.

From Eq.~(\ref{theorem_1:mean_inequality}), we have:
\begin{align}
\epsilon_j(t+1) &\le \left(1 - \dfrac{A_{j}}{\lambda t}\right) \epsilon_j(t) + \dfrac{A_{j-1}}{\lambda t}\epsilon_{j-1}(t) + Lt^{-1/2}\log t.\label{theorem_1:middle_eq}    
\end{align}
If we choose $t^{*} \ge t_1 \coloneqq \lceil \max_{j:j\le k}A_j/\lambda\rceil$, then $1 -A_j/(\lambda t) \ge 0$ for all $j \le k$ and $t \ge t^{*}$. Thus from the induction hypothesis, Eq.~(\ref{theorem_1:middle_eq}) leads to:
\begin{align}
\epsilon_j(t+1) &\le \left(1 - \dfrac{A_{j}}{\lambda t} + \dfrac{A_{j-1}}{\lambda t} \right)m^{*}\sqrt{t}\log t + Lt^{-1/2}\log t.\nonumber \\
&= m^{*}\sqrt{t}\log t + \left(L-A_jm^{*}/\lambda + A_{j-1}m^{*}/\lambda\right)t^{-1/2}\log t.\label{theorem_1:temp_1}  
\end{align}

Notice that, when $a \le 1/4$, the following inequality holds for all $t \ge 1$:
\begin{equation}
t^{1/2}\log t + at^{-1/2}\log t \le (t+1)^{1/2}\log (t+1).\label{eq:sup_inequality}
\end{equation}
Thus, when $m^{*} \ge 4L$, Eqs.~(\ref{theorem_1:temp_1}) and~(\ref{eq:sup_inequality}) lead to:
\begin{align} 
\epsilon_{j}(t+1) &\le m^{*}t^{1/2}\log t + \left(L-A_jm^{*}/\lambda + A_{j-1}m^{*}/\lambda\right)t^{-1/2}\log t \nonumber \\
&=  m^{*}\left(t^{1/2}\log t + \left(\dfrac{L}{m^{*}} - A_j/\lambda + A_{j-1}/\lambda\right)t^{-1/2}\log t \right) \nonumber \\
&\le m^{*}\sqrt{t+1}\log (t+1),
\end{align}
since $- A_j/\lambda + A_{j-1}/\lambda \le 0$.

By symmetry,  when $m^{*} \ge 4L$, one will also have $\epsilon_j(t+1) \ge -m^{*}\sqrt{t+1}\log (t+1)$ for all $j\le k$ and $t \ge t^{*}$. So the induction is complete if we choose $t^{*} = \max\{t_0,t_1\}$, with $t_1 = \lceil \max_{j:j\le k}A_j/\lambda\rceil$, and $m^{*} = \max\{4L,\left ((2t^{*} + 1)/\sqrt{t^{*}}\right)/\log t^{*}\}$. This completes the proof of {Eq.~(\ref{eq:mu_recursive}).}

{Equation~(\ref{eq:mu_recursive}) can be used to check that the $\mu_k$'s form a probability distribution. By summing up both sides of Eq.~(\ref{eq:mu_recursive}), we see that $\sum_k \mu_k = 1$. Rearranging Eq.~(\ref{eq:mu_recursive}), we have a recursive formula $\mu_k = A_{k-1}\mu_{k-1}/(\lambda+A_k)$, $k\ge1$ with $\mu_0 = \lambda/(\lambda + A_0)>0$, thus $\mu_k>0$ by induction for all $k\ge0$. This completes the proof of Lemma~\ref{theorem:mean_converge_in_degree}.}

\section{Additional details on PAFit-oneshot}
\label{sup_sec:binning}
\begin{figure*}[!h]
  \centering
  \includegraphics[width=\textwidth]{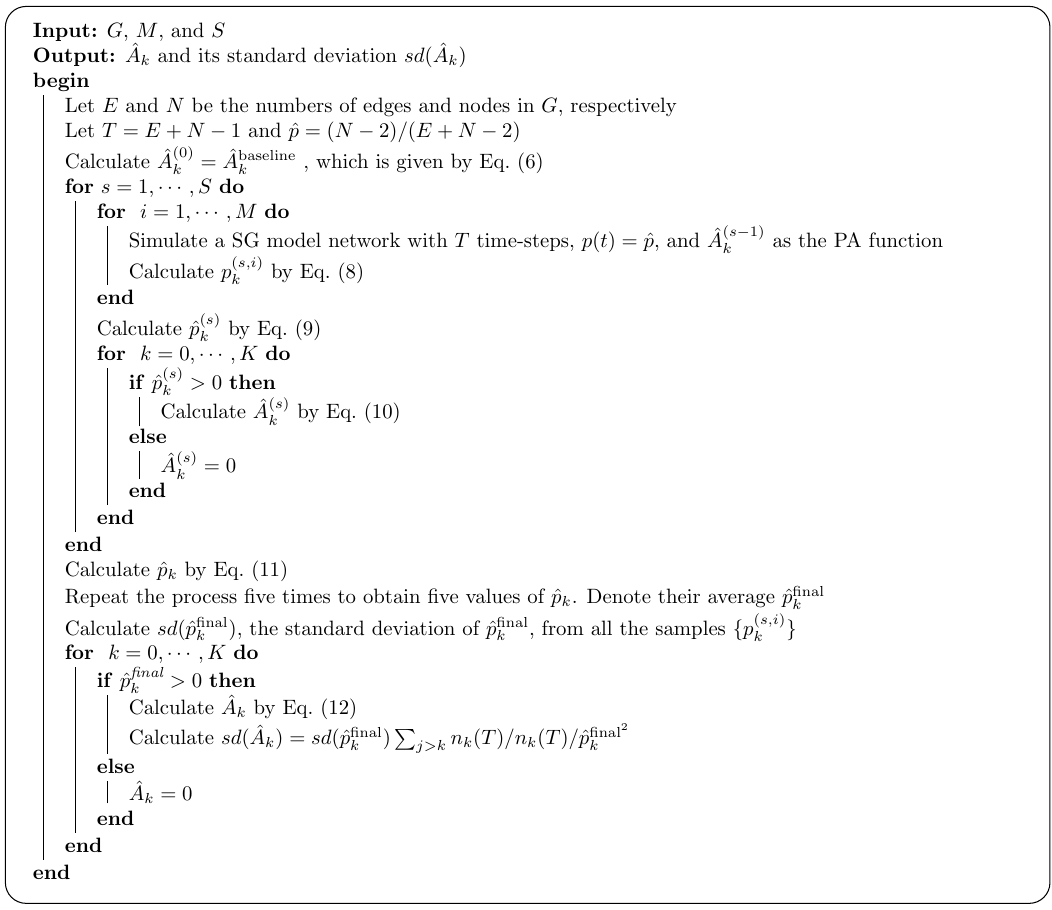}
  \caption{The workflow of PAFit-oneshot. \label{fig:mcPA-SI-fig}}
\end{figure*}

The {unbinned} version of PAFit-oneshot {is described} in Fig.~\ref{fig:mcPA-SI-fig}.

We discuss how to use PAFit-oneshot with binning. Suppose that we have $J$ bins $B_{1},\ldots,B_{J}$, each bin is a non-overlapping set of contiguous degrees, i.e., $B_g = \{k_{g},k_{g}+1,\ldots,k_{g}+l_g - 1\}$ with $l_g$ is the length of the $g$-th bin. Binning as a regularization technique works by enforcing the same PA value for all the degree $k$ in a bin, that is, $\theta_{g} = A_{k}$ for all $k \in B_{g}$. This regularization trades fine details of the PA function for a reduction in the number of parameters needed to be estimated.

For binning, we have the following theorem, whose proof is omitted since it is similar to the proof of Theorem~\ref{theorem:mean_converge}:
\begin{theorem}
Assume the same conditions as in Lemma~\ref{theorem:mean_converge_in_degree}. Suppose that $\theta_g = A_k$ for all $k \in B_g$. We have:
\begin{equation}
\theta_g = \lambda\dfrac{\mathbb{E}\sum_{k \in B_g}\sum_{j \ge k}n_k(t)}{\sum_{k \in B_g}\mathbb{E}n_k(t)} + o(1).\label{theorem:selective_binning}
\end{equation}
\end{theorem}
Plugging $t = T$ into Eq.~(\ref{theorem:selective_binning}) and correcting for the bias when we observe $G_T$, we obtain:
\begin{equation}
\theta_g \approx \dfrac{\sum_{k \in B_g}\sum_{j \ge k}n_k(T)}{\sum_{k_1 \in B_g:n_{k_1}(T) > 0}n_{k_1}(T)\mathbb{P}\left(n_{k_1}(T) > 0\right)}.
\end{equation}
At each round $s$, the estimate of $\theta_g$ is:
\begin{equation}
\hat{\theta}_{g}^{(s)} = \dfrac{\sum_{k \in B_g}\sum_{j > k}n_{k}(T)}{\sum_{k_1 \in B_g:n_{k_1}(T) > 0}n_{k_1}(T)\hat{p}_{k_1}^{(s)}}, \label{eq:mcPA_SI_bin_round_s}
\end{equation}
where $\hat{p}_{k}^{(s)}$ is given by Eq.~(\ref{eq:p_k_round_s}).
The final estimate of $\theta_g$ is:
\begin{equation}
\hat{\theta}_{g} = \dfrac{\sum_{k \in B_g}\sum_{j > k}n_{k}(T)}{\sum_{k_1 \in B_g:n_{k_1}(T) > 0}n_{k_1}(T)\hat{p}_{k_1}^{\text{final}}}. \label{eq:mcPA_SI_final_bin}
\end{equation}

\section{Maximum likelihood estimation when time-resolved data is available}
\label{sec:mle_full}
Here we present the MLE in the case when the growth process of the network is completely observed. This MLE serves as a gold standard for evaluating how well various methods perform in networks with time-resolved data.

We assume that we have time-resolved data, which means that we obtain $G_1,G_2,\ldots,G_T$. The log-likelihood function is:
\begin{equation}
l(\mathbf{A}) = \sum_{t=1}^{T-1}\sum_{k=0}^{K}m_k(t)\log{A_k} - \sum_{t = 1}^{T-1} m(t)\log \sum_{j=0}^{K}n_{j}(t) A_j, \label{eq: likelihood}
\end{equation}
with $\mathbf{A} = [A_0,A_1,\ldots,A_K]$ the parameter vector we need to estimate, $n_k(t)$ the number of nodes with degree $k$ at the onset of time-step $t$, and $m_k(t)$ the number of new edges connecting to a degree $k$ node at $t$. Here $K$ is the maximum degree to have appeared in the growth process.  We note that $\mathbf{A}$ is only identifiable up to a multiplicative constant, so in practice one often normalizes $\mathbf{A}$ so that $A_1 = 1$. When the scale of $\mathbf{A}$ is fixed, the log-likelihood function is known to be concave~\cite{pham2}. Solving the first-order optimality condition for this function leads to the following maximum likelihood equation:
\begin{equation}
A_k = \dfrac{\sum_{t=1}^{T-1}m_k(t)}{\sum_{t = 1}^{T-1} m(t) \dfrac{n_k(t)}{\sum_{j}n_{j}(t) A_j}}.\label{eq: likelihood-equation}
\end{equation}
Although a closed-form solution of this equation is unknown, the equation can be used as an update scheme that converges to the maximum likelihood estimator~\cite{pham2}. 

\section{Additional simulation results}
\label{sec_sup:sim}
\begin{figure*}[!h]
  \centering
  \includegraphics[width= \textwidth]{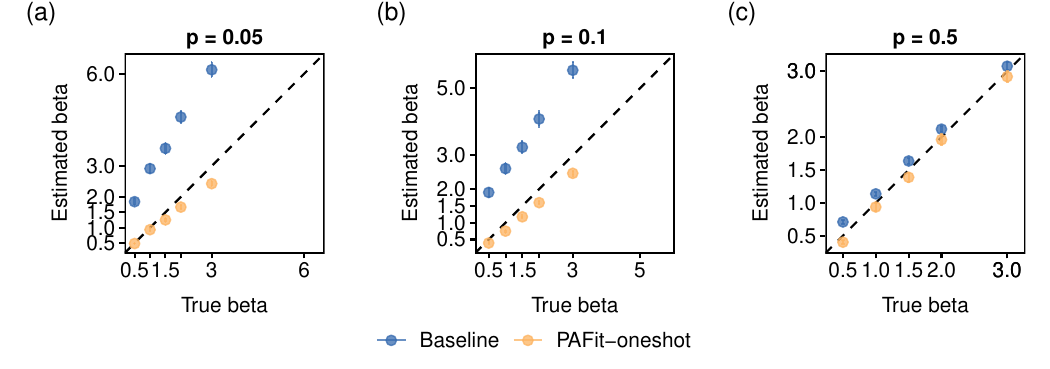}
  \caption{Simulation with $T = 5\times 10^5$ and $A_k = \max(k,1)/\left(1 + \beta \log\left(\max\left(k,1\right)\right)\right)$. For each case, $p(t) = p$, i.e., a constant. In each setting, the number of simulations is $50$. {(a) $p = 0.05$. (b) $p = 0.1$. (c) $p = 0.5$.}}\label{fig:simulation_redner}
\end{figure*}

\begin{figure*}[!h]
  \centering
  \includegraphics[width= \textwidth]{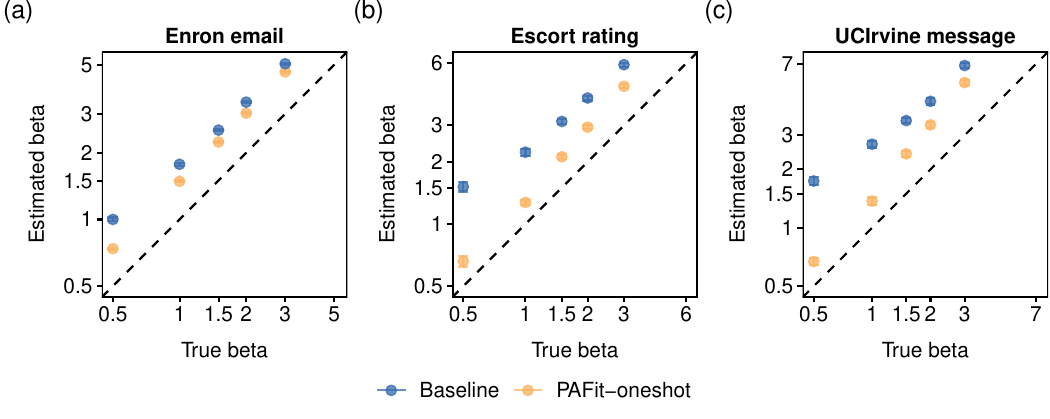}
  \caption{Simulation with real-world node-edge sequences obtained from three real-world networks and $A_k = \max(k,1)/\left(1 + \beta \log\left(\max\left(k,1\right)\right)\right)$. This PA form has been used in modelling citations in physics~\cite{use_Jeong3}. In each setting, the number of simulations is $50$. {(a) Enron email. (b) Escort rating. (c) UCIrvine message.}}\label{fig:simulation_p_t_real_redner}
\end{figure*}

Figures~\ref{fig:simulation_redner} and~\ref{fig:simulation_p_t_real_redner} show the result when we use $A_k = \max(k,1)/\left(1 + \beta \log\left(\max\left(k,1\right)\right)\right)$ as the true PA function for the case when $p(t)$ is held fixed and the case when $p(t)$ is varied based on real-world data. This functional form has been used in modelling citations in physics~\cite{use_Jeong3}. The results show that the proposed method outperformed the baseline with this PA function too.

\begin{figure*}[!h]
  \centering
  \includegraphics[width= \textwidth]{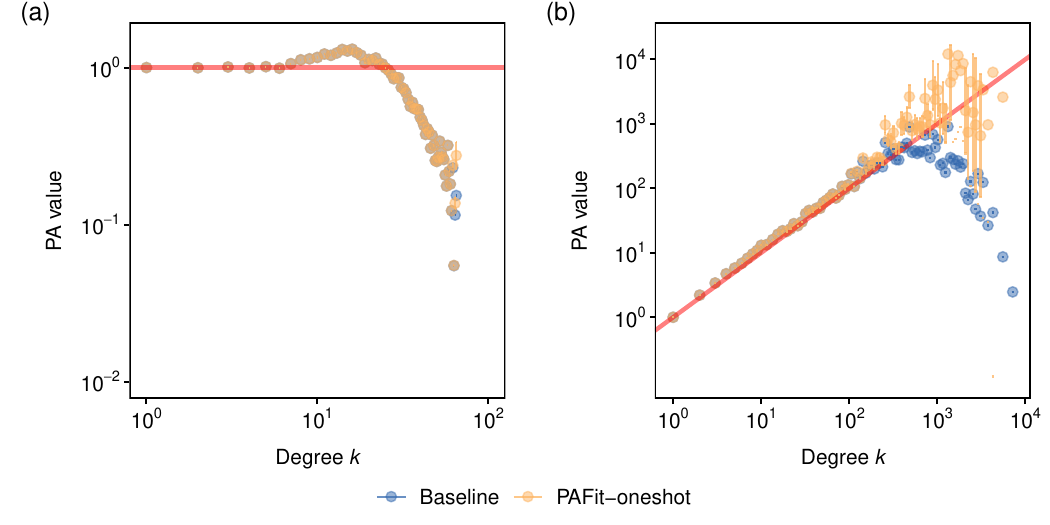}
  \caption{Two typical runs in simulations with Enron email network when we use $A_k = \max (k,1)^{\alpha}$. {(a)} $A_k = 1$ ($\alpha = 0$). PAFit-oneshot failed to remove the waterfall artefact. {(b)} $A_k = k$ ($\alpha = 1$). PAFit-oneshot corrected the waterfall artefact and gave good estimation values.\label{fig:one_example_enron}}
\end{figure*}

For the results in Fig.~\ref{fig:simulation_p_t_real_log_linear}, Fig.~\ref{fig:one_example_enron} shows two runs that are typical for the situations when $\alpha$ is small and when $\alpha$ is large, respectively. 

\section{Additional information for real-world networks with time-resolved data}\label{sec_sup:validating}
Table~\ref{sup_tab: dataset_full_timeseries} shows several statistics for the networks in Fig.~\ref{fig:validating_real_PA}. 
\begin{table*}[!h]
\caption{Summary statistics of networks with time-resolved data. $N$ and $E$ is the number of nodes and edges, respectively. $T$ is the number of time-steps. The quantity $\bar{p}$ is the average ratio of nodes. The quantity $d_{max}$ is the maximum degree in the network.} \label{sup_tab: dataset_full_timeseries}
\centering
\begin{tabularx}{\textwidth}{l l l l l l}
\hline
\textbf{Network} &$N$ &$E$ &  $T = N + E - 1$ & $\bar{p} = N/T$ & $d_{max}$\\ \hline
Enron email&  87272 & 1148072 & 1235343 & 0.07& 5419\\
Sexual escort rating & 6624 & 50632 & 57255 & 0.12 & 615\\
UCIrvine forum message &1899 & 59835	 & 61733 & 0.03 & 1546\\
\hline
\end{tabularx}
\end{table*}

\begin{figure*}[!h]
  \centering
  \includegraphics[width= \textwidth]{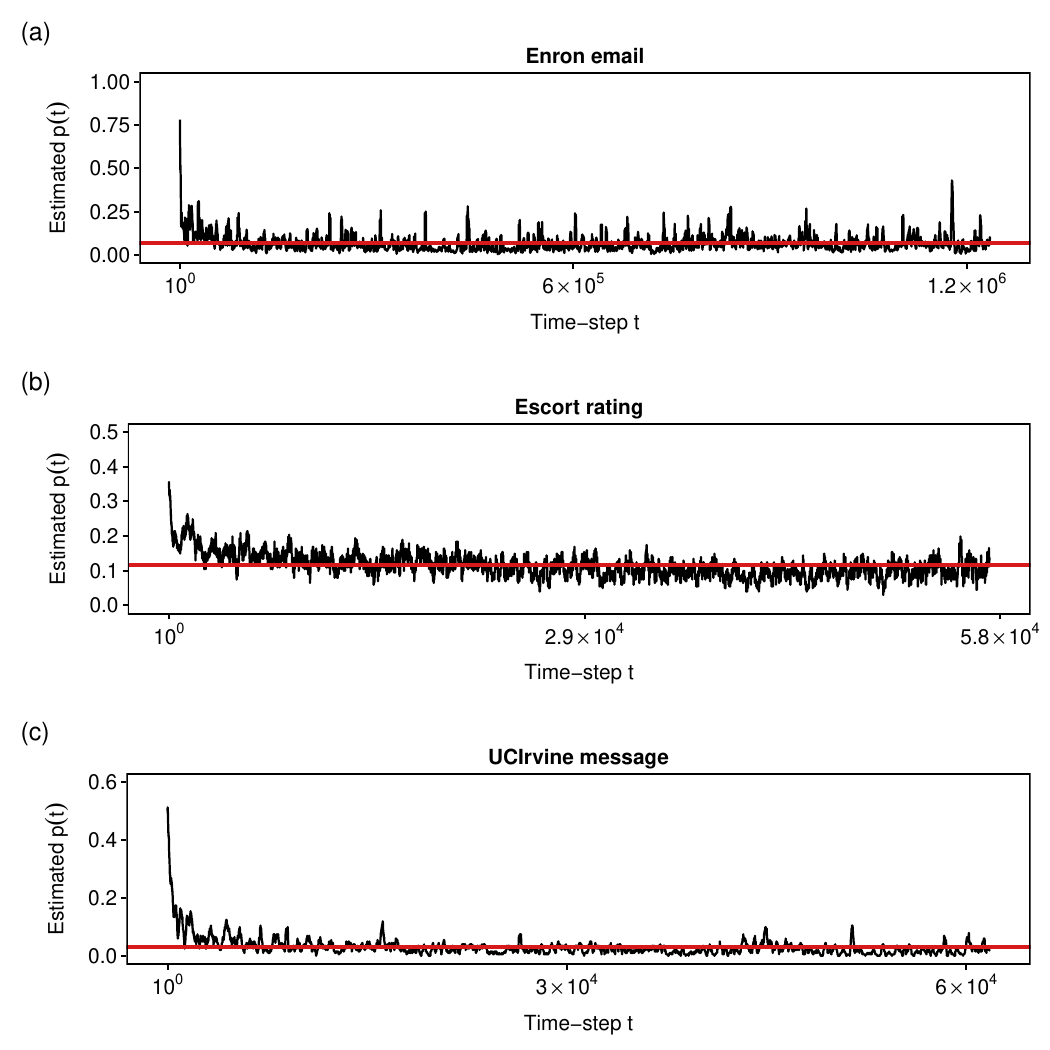}
  \caption{Estimated $p(t)$ in three real-world networks with time-resolved data: Enron email, the Escort rating, and the UCIrvine forum message networks.  At each time-step $t$, we estimate $p(t)$ by the average of the number of time-steps that contains a new node in the range $t \pm 500$ in the Enron email network and $t \pm 50$ in the Sexual escort rating and the UCIrvine forum message networks. The horizontal red lines indicate the average $\bar{p}$ in the networks, whose values can be found in Table~\ref{sup_tab: dataset_full_timeseries}. {(a) Enron email. (b) Escort rating. (c) UCIrvine message.}}\label{sup_fig:real_p}
\end{figure*}

Figure~\ref{sup_fig:real_p} show the estimated $p(t)$ in the three real-world networks.
\clearpage
\section{Additional information for real-world networks without time-resolved data}\label{sec_sup:unmeasurable}
\begin{figure*}[!h]
  \centering
  \includegraphics[width= \textwidth]{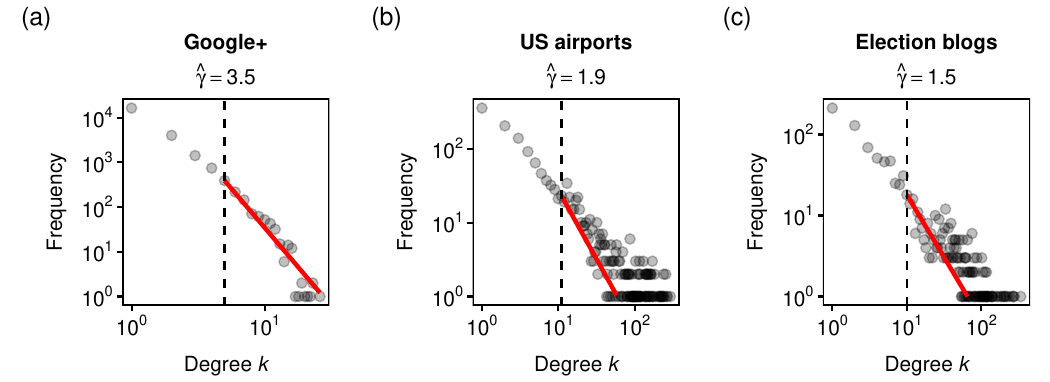}
  \caption{Degree distributions of real-world networks with no time-step information. The power-law exponent $\gamma$ is estimated by maximum likelihood estimation. The degree threshold $k_{min}$ from which the power-law property holds is estimated by Clauset method in the Google+ and US airport networks. They are $k_{min} = 11$ and $k_{min} = 5$, respectively. In the Election blog network, $k_{min} = 10$ is chosen by visual inspection. {(a) Google+. (b) US airports. (c) Election blogs.}}\label{sup_fig:unmeasurable_deg}
\end{figure*}

Table~\ref{sup_tab: dataset_no_timeseries_stats} shows several statistics for the networks in Fig.~\ref{fig:unmeasurable_real_PA}. The degree distributions are given in Fig.~\ref{sup_fig:unmeasurable_deg}. The scale-free form is fitted to the empirical degree distribution by method of Clauset~\cite{clauset}.

\begin{table*}[!h]
\caption{Several statistics for the real-world networks with no time-step information. $N$ and $E$ is the number of nodes and edges, respectively. {The quantity} $\bar{p}$ is the average ratio of nodes. {The quantity} $d_{max}$ is the maximum degree in the network.\label{sup_tab: dataset_no_timeseries_stats}}
\centering
\begin{tabularx}{\textwidth}{ l l l l l}
\hline
\textbf{Network} &$N$ &$E$  & $\bar{p} = N/(N + E)$ & $d_{max}$\\ \hline
US Airports & 6624 & 50632  & 0.12 & 615\\ 
Google+ & 23628 & 39242 & 0.38 & 26\\
2004 US Election blogs &1224 & 19025  & 0.06 & 467 \\
\hline
\end{tabularx}
\end{table*}

\begin{thebibliography}{00}

\bibitem{us_blog}
Adamic, L.~A. {\&} Glance, N. (2005)  The political blogosphere and the 2004
  U.S. election: divided they blog. In {\em Proceedings of the 3rd
  International Workshop on Link Discovery}, LinkKDD '05, page 36–43, New
  York, NY, USA. Association for Computing Machinery.

\bibitem{barabasi-albert}
Albert, R. {\&} Barab{\'a}si, A. (1999)  Emergence of scaling in random networks. {\em Science}, \textbf{286}, 509--512.

\bibitem{PA_edge_step}
Alves, C., Ribeiro, R. {\&} Sanchis, R. (2021)  Preferential attachment random
  graphs with edge-step functions. {\em Journal of Theoretical Probability},
  \textbf{34}(1), 438--476.

\bibitem{timeline_mcmc_icml}
Bez\'{a}kov\'{a}, I., Kalai, A. {\&} Santhanam, R. (2006)  Graph model
  selection using maximum likelihood. In {\em Proceedings of the 23rd
  International Conference on Machine Learning}, ICML '06, pages 105--112, New
  York, NY, USA. ACM.


\bibitem{bollobas_directed_graph}
Bollob\'{a}s, B., Borgs, C., Chayes, J. {\&} Riordan, O. (2003)  Directed
  scale-free graphs. In {\em Proceedings of the Fourteenth Annual ACM-SIAM
  Symposium on Discrete Algorithms}, SODA '03, pages 132--139, Philadelphia,
  PA, USA. Society for Industrial and Applied Mathematics.

\bibitem{bollobas_2001}
Bollob\'{a}s, B., Riordan, O., Spencer, J. {\&} Tusn\'{a}dy, G. (2001)  The
  degree sequence of a scale-free random graph process. {\em Random Structures
  \& Algorithms}, \textbf{18}(3), 279--290.

\bibitem{bradford_law}
Bradford, S.~C. (1985)  Sources of information on specific subjects. {\em J.
  Inf. Sci.}, \textbf{10}(4), 173–180.

\bibitem{cantwell_2021}
Cantwell, G.~T., St-Onge, G. {\&} Young, J.-G. (2021)  Inference, model
  selection, and the combinatorics of growing trees. {\em Phys. Rev. Lett.},
  \textbf{126}, 038301.

\bibitem{clauset}
Clauset, A., Shalizi, C.~R. {\&} Newman, M. E.~J. (2009)  Power-law
  distributions in empirical data. {\em SIAM Review}, \textbf{51}(4), 661--703.

\bibitem{cooper_model}
Cooper, C. {\&} Frieze, A. (2003)  A general model of web graphs. {\em Random
  Structures \& Algorithms}, \textbf{22}(3), 311--335.

\bibitem{deijfen2009}
Deijfen, M., van~den Esker, H., van~der Hofstad, R. {\&} Hooghiemstra, G.
  (2009)  {A preferential attachment model with random initial degrees}. {\em
  Arkiv för Matematik}, \textbf{47}(1), 41 -- 72.

\bibitem{dereich2009}
Dereich, S. {\&} Mörters, P. (2009)  Random networks with sublinear
  preferential attachment: Degree evolutions. {\em Electron. J. Probab.},
  \textbf{14}, 1222--1267.
  
\bibitem{feng_2020}
Feng, M., Deng, L.-J., Chen, F., Perc, M. {\&} Kurths, J. (2020)  The
  accumulative law and its probability model: an extension of the Pareto
  distribution and the log-normal distribution. {\em Proceedings of the Royal
  Society A: Mathematical, Physical and Engineering Sciences},
  \textbf{476}(2237), 20200019.
  
\bibitem{feng_2017}
Feng, M., Qu, H., Yi, Z. {\&} Kurths, J. (2018)  Subnormal distribution derived
  from evolving networks with variable elements. {\em IEEE Transactions on
  Cybernetics}, \textbf{48}(9), 2556--2568.

\bibitem{gao_2016}
Gao, F. {\&} van~der Vaart, A. (2017)  On the asymptotic normality of
  estimating the affine preferential attachment network models with random
  initial degrees. {\em Stochastic Processes and their Applications},
  \textbf{127}(11), 3754--3775.

\bibitem{gao2017}
Gao, F., van~der Vaart, A., Castro, R. {\&} van~der Hofstad, R. (2017)
  Consistent estimation in general sublinear preferential attachment trees.
  {\em Electron. J. Statist.}, \textbf{11}(2), 3979--3999.

\bibitem{Gomez}
G\'{o}mez, V., Kappen, H.~J. {\&} Kaltenbrunner, A. (2011)  Modeling the
  structure and evolution of discussion cascades. In {\em Proceedings of the
  22Nd ACM Conference on Hypertext and Hypermedia}, HT '11, pages 181--190, New
  York, NY, USA. ACM.

\bibitem{timeline_adaptive_sampling}
Guetz, A.~N. {\&} Holmes, S.~P. (2011)  Adaptive importance sampling for
  network growth models. {\em Annals of Operations Research}, \textbf{189}(1),
  187--203.

\bibitem{non_decreasing_PA_example}
Hagberg, O. {\&} Wiuf, C. (2006)  Convergence properties of the degree
  distribution of some growing network models. {\em Bulletin of Mathematical
  Biology}, \textbf{68}(6), 1275.

\bibitem{jeong}
Jeong, H., N{\'e}da, Z. {\&} Barab{\'a}si, A. (2003)  Measuring preferential
  attachment in evolving networks. {\em Europhysics Letters}, \textbf{61}(61),
  567--572.

\bibitem{enron}
Klimt, B. {\&} Yang, Y. (2004)  The Enron corpus: A new dataset for email
  classification research. In {\em In Proc. European Conf. on Machine
  Learning}, pages 217--226.

\bibitem{krapi2}
Krapivsky, P., Rodgers, G. {\&} Redner, S. (2001a)  Degree distributions of
  growing networks. {\em Physical Review Letters}, \textbf{86}(23), 5401--5404.

\bibitem{krapi}
Krapivsky, P., Rodgers, G. {\&} Redner, S. (2001b)  Organization of growing
  networks. {\em Physical Review E}, page 066123.

\bibitem{gplus}
Leskovec, J. {\&} Mcauley, J. (2012)  Learning to discover social circles in
  ego networks. In Pereira, F., Burges, C. J.~C., Bottou, L. {\&} Weinberger,
  K.~Q., editors, {\em Advances in Neural Information Processing Systems},
  volume~25, pages 539--547. Curran Associates, Inc.

\bibitem{power_law_myth}
Lima-Mendez, G. {\&} van Helden, J. (2009)  The powerful law of the power law
  and other myths in network biology. {\em Mol. BioSyst.}, \textbf{5},
  1482--1493.

\bibitem{lotka_law}
Lotka, A.~J. (1926)  The frequency distribution of scientific productivity.
  {\em Journal of the Washington Academy of Sciences}, \textbf{16}(12),
  317--323.

\bibitem{link_prediction_survey}
Lü, L. {\&} Zhou, T. (2011)  Link prediction in complex networks: A survey.
  {\em Physica A: Statistical Mechanics and its Applications}, \textbf{390}(6),
  1150--1170.

\bibitem{massen}
Massen, C. {\&} Jonathan, P. (2007)  Preferential attachment during the
  evolution of a potential energy landscape. {\em The Journal of Chemical
  Physics}, \textbf{127}, 114306.

\bibitem{Merton56}
Merton, R.~K. (1968)  The Matthew effect in science. {\em Science},
  \textbf{159}(3810), 56--63.

\bibitem{newman2001clustering}
Newman, M. (2001)  Clustering and preferential attachment in growing networks.
  {\em Physical Review E}, \textbf{64}(2), 025102.

\bibitem{tore_usairport}
Opsahl, T. (2011)  {W}hy anchorage is not (that) important: binary ties and sample selection.
  \url{https://toreopsahl.com/2011/08/12/why-anchorage-is-not-that-important-binary-ties-and-sample-selection/}.

\bibitem{ucirvine}
Opsahl, T. {\&} Panzarasa, P. (2009)  Clustering in weighted networks. {\em
  Social Networks}, \textbf{31}, 155--163.

\bibitem{perc}
Perc, M. (2014)  The Matthew effect in empirical data. {\em Journal of The
  Royal Society Interface}, \textbf{11}(98).

\bibitem{pham2}
Pham, T., Sheridan, P. {\&} Shimodaira, H. (2015)  {PAFit}: {a} statistical
  method for measuring preferential attachment in temporal complex
  networks. {\em PLOS ONE}, \textbf{10}(9), e0137796.

\bibitem{pham3}
Pham, T., Sheridan, P. {\&} Shimodaira, H. (2016)  Joint estimation of
  preferential attachment and node fitness in growing complex
  networks. {\em Scientific Reports}, \textbf{6}.

\bibitem{pham_jss}
Pham, T., Sheridan, P. {\&} Shimodaira, H. (2020)  {PAFit}: An {R} package for
  the non-parametric estimation of preferential attachment and node fitness in
  temporal complex networks. {\em Journal of Statistical Software, Articles},
  \textbf{92}(3), 1--30.

\bibitem{price2}
Price, D. d.~S. (1965)  Networks of scientific papers. {\em Science},
  \textbf{149}(3683), 510--515.

\bibitem{price1}
Price, D. d.~S. (1976)  A general theory of bibliometric and other cumulative
  advantage processes. {\em Journal of the American Society for Information
  Science}, \textbf{27}, 292--306.

\bibitem{use_Jeong3}
Redner, S. (2005)  Citation statistics from 110 years of physical review. {\em
  Physics Today}, \textbf{58 (6)}, 49--54.

\bibitem{escort}
Rocha, L. E.~C., Liljeros, F. {\&} Holme, P. (2010)  Information dynamics shape
  the sexual networks of Internet-mediated prostitution. {\em Proceedings of
  the National Academy of Sciences}, \textbf{107}(13), 5706--5711.

\bibitem{herding_1}
Rodgers, G. {\&} Zheng, D. (2002)  A herding model with preferential attachment
  and fragmentation. {\em Physica A: Statistical Mechanics and its
  Applications}, \textbf{308}(1), 375--380.

\bibitem{rudas_2007}
Rudas, A., T{\'o}th, B. {\&} Valk{\'o}, B. (2007)  Random trees and general
  branching processes. {\em Random Structures \& Algorithms}, \textbf{31}(2),
  186--202.

\bibitem{power_law_finite_size_effect}
Serafino, M., Cimini, G., Maritan, A., Rinaldo, A., Suweis, S., Banavar, J.~R.
  {\&} Caldarelli, G. (2021)  True scale-free networks hidden by finite size
  effects. {\em Proceedings of the National Academy of Sciences},
  \textbf{118}(2).

\bibitem{sheridan_2008}
Sheridan, P., Yagahara, Y. {\&} Shimodaira, H. (2008)  A preferential
  attachment model with Poisson growth for scale-free networks. {\em Annals of
  the Institute of Statistical Mathematics}, \textbf{60}(4), 747--761.

\bibitem{sheridan}
Sheridan, P., Yagahara, Y. {\&} Shimodaira, H. (2012)  {Measuring preferential
  attachment in growing networks with missing-timelines using Markov chain
  Monte Carlo}. {\em Physica A Statistical Mechanics and its Applications},
  \textbf{391}, 5031--5040.

\bibitem{simon}
Simon, H.~A. (1955)  On a class of skew distribution functions. {\em
  Biometrika}, \textbf{42}(3-4), 425--440.

\bibitem{Taylor7629}
Taylor, J. {\&} Tibshirani, R.~J. (2015)  Statistical learning and selective
  inference. {\em Proceedings of the National Academy of Sciences},
  \textbf{112}(25), 7629--7634.

\bibitem{tibshirani2016exactpsi}
Tibshirani, R.~J., Taylor, J., Lockhart, R. {\&} Tibshirani, R. (2016)  Exact
  Post-Selection Inference for Sequential Regression Procedures. {\em Journal
  of the American Statistical Association}, \textbf{111}(514), 600--620.

\bibitem{clauset_powerlaw_2}
Virkar, Y. {\&} Clauset, A. (2014)  Power-law distributions in binned empirical
  data. {\em The Annals of Applied Statistics}, \textbf{8}(1), 89--119.

\bibitem{wan2017}
Wan, P., Wang, T., Davis, R.~A. {\&} Resnick, S.~I. (2017)  Fitting the linear
  preferential attachment model. {\em Electron. J. Statist.}, \textbf{11}(2),
  3738--3780.

\bibitem{yule}
Yule, G.~U. (1925)  A mathematical theory of evolution, based on the
  conclusions of {D}r. {J}.{C}. {W}illis,{F}.{R}.{S}.. {\em Philosophical
  Transactions of the Royal Society of London B: Biological Sciences},
  \textbf{213}(402-410), 21--87.

\end{thebibliography}

\end{document}